\documentclass[10pt]{article}

\usepackage{amsmath}
\usepackage{amsfonts}
\usepackage{amssymb}
\usepackage{graphicx}
\usepackage{authblk}

\usepackage{makeidx}
\usepackage{multicol}
\usepackage{rotating}
\usepackage{multirow}
\usepackage{amssymb}
\usepackage{ifthen}
\usepackage[dvipsnames]{xcolor}
\usepackage{boxedminipage}
\usepackage{fancyvrb}
\usepackage[dotinlabels]{titletoc}
\usepackage{colortbl}
\usepackage{array}
\usepackage{wrapfig}
\usepackage{moreverb}
\usepackage[Lenny]{fncychap}
\usepackage{tikz}
\usetikzlibrary{arrows,matrix,decorations.pathreplacing,shapes,backgrounds}
\usepackage{textcomp}
\usepackage{pifont}
\usepackage{hyperref}
\usepackage[totoc]{idxlayout}
\usepackage{dirtree}
\usepackage{listings}
\usepackage{makecell}
\renewcommand{\hline}{\Xhline{2\arrayrulewidth}}

\usepackage{datetime}

\definecolor{mygreen}{rgb}{0,0.6,0}
\definecolor{mygray}{rgb}{0.5,0.5,0.5}
\definecolor{mymauve}{rgb}{0.58,0,0.82}

\renewcommand{\em}{\it}   

\newcolumntype{I}{!{\vrule width 1.5pt}}
\newlength\savedwidth
\newcommand\whline{\noalign{\global\savedwidth\arrayrulewidth
                            \global\arrayrulewidth 1.5pt}%
           \hline
           \noalign{\global\arrayrulewidth\savedwidth}}

\newcommand{\FlaTwoByOne}[2]{
\left(
\begin{array}{c}
#1 \\ \hline
#2
\end{array}
\right)
}

\newcommand{\FlaTwoByOneSingleLine}[2]{
\left(
\begin{array}{c}
#1 \\ 
#2
\end{array}
\right)
}

\newcommand{\FlaThreeByOneT}[3]{
\left(
\begin{array}{c}
#1 \\ 
#2 \\ \hline
#3
\end{array}
\right)
}

\newcommand{\FlaThreeByOneB}[3]{
\left(
\begin{array}{c}
#1 \\ \hline
#2 \\ 
#3
\end{array}
\right)
}

\newcommand{\operation}{}

\newcommand{\routinename}{}

\newcommand{\precondition}{~}

\newcommand{\postcondition}{~}

\newcommand{\invariant}{~}

\newcommand{\guard}{~}

\newcommand{\partitionings}{~}

\newcommand{\partitionsizes}{~}

\newcommand{\blocksize}{blank}

\newcommand{\repartitionings}{~}

\newcommand{\repartitionsizes}{~}

\newcommand{\moveboundaries}{~}

\newcommand{\beforeupdate}{~}

\newcommand{\afterupdate}{~}

\newcommand{\update}{~}

\newcommand{\NoShow}[1]{}

\newcommand{\FlaAlgorithm}{
\begin{tabular}{|l|} \hline
$\mbox{\color{blue}Algorithm:~}\routinename$
\\ \hline
\partitionings 
\\
$\mbox{\color{blue} ~~~where~}$ \partitionsizes 
\\ 
$\mbox{\color{blue}while~} \ShowGuard \mbox{~\color{blue} do}$
\\
\ifthenelse{\equal{\blocksize}{1}}{}%
{%
\ifthenelse{ \equal{\blocksize}{blank} }{}%
{~~~~{\bf Determine block size $ \blocksize $}\\}%
}
~~~~ 
\repartitionings 
\NoShow{\\
~~~$\mbox{\color{blue} ~~~where~}$ \repartitionsizes
}
\\ \hline
~~~~  \update 
\\ \hline
~~~~ 
\moveboundaries 
\\
$\mbox{\color{blue} endwhile} $
\\ \hline 
\end{tabular}
}

\newcounter{WSStep}
\newcommand{\ShowPrecondition}{\ifthenelse{\value{WSStep}<1}%
   {{\color{white} \precondition}}
   {\ifthenelse{\value{WSStep}=1}%
    {\color{red} \precondition}
    {\color{black} \precondition}}}
\newcommand{\ShowPostcondition}{\ifthenelse{\value{WSStep}<1}%
   {{\color{white} \postcondition}}
   {\ifthenelse{\value{WSStep}=1}%
    {\color{red} \postcondition}
    {\color{black} \postcondition}}}
\newcommand{\ShowInvariant}{\ifthenelse{\value{WSStep}<2}%
   {{\color{white} \invariant}}
   {\ifthenelse{\value{WSStep}=2}%
    {\color{red} \invariant}
    {\color{black} \invariant}}}
\newcommand{\ShowGuard}{\ifthenelse{\value{WSStep}<3}%
   {{\color{lightgray!25} \guard}}
   {\ifthenelse{\value{WSStep}=3}%
    {\color{red} \guard}
    {\color{black} \guard}}}
\newcommand{\ShowGuardTwo}{\ifthenelse{\value{WSStep}<3}%
   {{\color{white} \guard}}
   {\ifthenelse{\value{WSStep}=3}%
    {\color{red} \guard}
    {\color{black} \guard}}}
\newcommand{\ShowPartitionings}{\ifthenelse{\value{WSStep}<4}%
   {{\color{lightgray!25} \partitionings}}%
   {\ifthenelse{\value{WSStep}=4}%
    {\color{red} \partitionings}%
    {\color{black} \partitionings}}}
\newcommand{\ShowPartitionSizes}{\ifthenelse{\value{WSStep}<4}%
   {{\color{lightgray!25} \partitionsizes}}
   {\ifthenelse{\value{WSStep}=4}%
    {\color{red} \partitionsizes}
    {\color{black} \partitionsizes}}}
\newcommand{\ShowRepartitionings}{\ifthenelse{\value{WSStep}<5}%
   {{\color{lightgray!25} \repartitionings}}
   {\ifthenelse{\value{WSStep}=5}%
    {\color{red} \repartitionings}
    {\color{black} \repartitionings}}}
\newcommand{\ShowRepartitionSizes}{\ifthenelse{\value{WSStep}<5}%
   {{\color{lightgray!25} \repartitionsizes}}
   {\ifthenelse{\value{WSStep}=5}%
    {\color{red} \repartitionsizes}
    {\color{black} \repartitionsizes}}}
\newcommand{\ShowMoveBoundaries}{\ifthenelse{\value{WSStep}<5}%
   {{\color{lightgray!25} \moveboundaries}}
   {\ifthenelse{\value{WSStep}=5}%
    {\color{red} \moveboundaries}
    {\color{black} \moveboundaries}}}
\newcommand{\ShowBeforeUpdate}{\ifthenelse{\value{WSStep}<6}%
   {{\color{white} \beforeupdate}}
   {\ifthenelse{\value{WSStep}=6}%
    {\color{red} \beforeupdate}
    {\color{black} \beforeupdate}}}
\newcommand{\ShowAfterUpdate}{\ifthenelse{\value{WSStep}<7}%
   {{\color{white} \afterupdate}}
   {\ifthenelse{\value{WSStep}=7}%
    {\color{red} \afterupdate}
    {\color{black} \afterupdate}}}
\newcommand{\ShowUpdate}{\ifthenelse{\value{WSStep}<8}%
   {{\color{lightgray!25} \update}}
   {\ifthenelse{\value{WSStep}=8}%
    {\color{red} \update}
    {\color{black} \update}}}

\newcommand{\FlaWorksheet}{
\begin{tabular}{| c | p{0.9\textwidth} |}\hline
Step & $\mbox{\color{blue}Algorithm:~}\routinename$
\\ \hline
1a &%
$ \left\{ 
\begin{minipage}{0.88\textwidth} 
$\ShowPrecondition$  
\end{minipage}
\right\}
$%
\\ \hline
\rowcolor{lightgray!25}   
4 & %
\begin{minipage}{0.88\textwidth}%
\vspace{0.05in}
\ShowPartitionings~ \\
\mbox{\color{blue} ~~~where~} \ShowPartitionSizes
\end{minipage}
\\ \hline
2 & 
$ \left\{ 
\begin{minipage}{0.88\textwidth} 
$\ShowInvariant $
\end{minipage}
\right\} $ 
\\ \hline
\rowcolor{lightgray!25}   
3 &$\mbox{\color{blue}while~} \ShowGuard \mbox{~\color{blue} do}$
\\ \hline 
2,3 &  
$
\left\{
\begin{minipage}[t]{0.88\textwidth}%
$
~~~~ \ShowInvariant 
\wedge \ShowGuardTwo$
\end{minipage}
\right\}
$ 
\\ \hline
\rowcolor{lightgray!25}   
5a & ~~~~ \begin{minipage}{0.85\textwidth}%
\vspace{0.05in}
\ifthenelse{\equal{\blocksize}{1}}{}%
{%
\ifthenelse{ \equal{\blocksize}{blank} }{}%
{{\bf Determine block size $ \blocksize $}\\}%
}
\ShowRepartitionings~ 
\NoShow{\\
$\mbox{\color{blue} ~~~where~}$ \ShowRepartitionSizes}
\end{minipage}
\\ \hline
6 & 
$ \left\{ 
\begin{minipage}{0.88\textwidth} 
~~~~ \ShowBeforeUpdate 
\end{minipage}
\right\}
$
\\ \hline
\rowcolor{lightgray!25}  
8 & ~~~~  \ShowUpdate 
\\ \hline 
7 & 
$ \left\{ 
\begin{minipage}{0.88\textwidth} 
~~~~ \ShowAfterUpdate 
\end{minipage}
\right\}
$
\\ \hline
\rowcolor{lightgray!25}   
5b & ~~~~ \begin{minipage}{0.85\textwidth}%
\vspace{0.05in}
\ShowMoveBoundaries~
\end{minipage}
\\ \hline
2 & 
$ \left\{ 
\begin{minipage}{0.88\textwidth} 
~~~~ $ \ShowInvariant  $ 
\end{minipage}
\right\}
$
\\ \hline
\rowcolor{lightgray!25}  
 &$\mbox{\color{blue} endwhile} $
\\ \hline 
2,3 & 
$ \left\{ 
\begin{minipage}{0.88\textwidth} 
$ \ShowInvariant \wedge \neg( \ShowGuardTwo )$ 
\end{minipage}
\right\}
$
\\ \hline
1b & 
$ \left\{ 
\begin{minipage}{0.88\textwidth} 
$ \ShowPostcondition $ 
\end{minipage}
\right\}
$
\\ \hline
\end{tabular}
}

\newcommand{\FlaWorksheetNine}{
\begin{tabular}{| c | p{0.9\textwidth} |}\hline
{\color{white}Step} & $\mbox{\color{blue}Algorithm:~}\routinename$
\\ \hline
 &%
$ \phantom{\left\{ 
\begin{minipage}{0.88\textwidth} 
$\ShowPrecondition$  
\end{minipage}
\right\}}
$%
\\ \hline
\rowcolor{lightgray!25}   
& %
\begin{minipage}{0.88\textwidth}%
\vspace{0.05in}
\ShowPartitionings~ \\
\mbox{\color{blue} ~~~where~} \ShowPartitionSizes
\end{minipage}
\\ \hline
& 
$ \phantom{\left\{ 
\begin{minipage}{0.88\textwidth} 
$\ShowInvariant $
\end{minipage}
\right\}} $ 
\\ \hline
\rowcolor{lightgray!25}   
&$\mbox{\color{blue}while~} \ShowGuard \mbox{~\color{blue} do}$
\\ \hline 
 &  
$
\phantom{\left\{
\begin{minipage}[t]{0.88\textwidth}%
~~~~$
\ShowInvariant 
\wedge \ShowGuardTwo$
\end{minipage}
\right\}}
$ 
\\ \hline
\rowcolor{lightgray!25}   
 & ~~~~ \begin{minipage}{0.85\textwidth}%
\vspace{0.05in}
\ifthenelse{\equal{\blocksize}{1}}{}%
{%
\ifthenelse{ \equal{\blocksize}{blank} }{}%
{{\bf Determine block size $ \blocksize $}\\}%
}
\ShowRepartitionings~ \\
$\mbox{\color{blue} ~~~where~}$ \ShowRepartitionSizes
\end{minipage}
\\ \hline
& 
$ \phantom{\left\{ 
\begin{minipage}{0.88\textwidth} 
~~~~ \ShowBeforeUpdate 
\end{minipage}
\right\}}
$
\\ \hline
\rowcolor{lightgray!25}  
 & ~~~~  \ShowUpdate 
\\ \hline 
& 
$ \phantom{\left\{ 
\begin{minipage}{0.88\textwidth} 
~~~~ \ShowAfterUpdate 
\end{minipage}
\right\}}
$
\\ \hline
\rowcolor{lightgray!25}   
 & ~~~~ \begin{minipage}{0.85\textwidth}%
\vspace{0.05in}
\ShowMoveBoundaries~
\end{minipage}
\\ \hline
& 
$ \phantom{\left\{ 
\begin{minipage}{0.88\textwidth} 
~~~~ $ \ShowInvariant  $ 
\end{minipage}
\right\}}
$
\\ \hline
\rowcolor{lightgray!25}  
 &$\mbox{\color{blue} endwhile} $
\\ \hline 
& 
$ \phantom{\left\{ 
\begin{minipage}{0.88\textwidth} 
$ \ShowInvariant \wedge \neg( \ShowGuardTwo )$ 
\end{minipage}
\right\}}
$
\\ \hline
& 
$ \phantom{\left\{ 
\begin{minipage}{0.88\textwidth} 
$ \ShowPostcondition $ 
\end{minipage}
\right\}}
$
\\ \hline
\end{tabular}
}

\newcommand{\PseudocodeAssert}[1]{%
$%
\left\{ 
\colorbox{white}{
	\begin{minipage}{0.9\textwidth}
#1 
\end{minipage}%
}
\right\}  
$
\\ \hline 
}

\newcommand{\PseudocodeAssertIndent}[1]{%
\cellcolor{white}%
\mbox{~~~~} %
$%
\left\{ 
\cellcolor{white}{
	\begin{minipage}{0.9\textwidth}
#1
\end{minipage}
}
\right\}  
$
\\ \hline 
}

\newcommand{\PseudocodeStatement}[1]{
	\rowcolor{lightgray!25}
        \begin{tabular}{@{}l}
	#1
        \end{tabular}
	\\ \hline
}

\newcommand{\PseudocodeStatementIndent}[1]{
	\rowcolor{lightgray!25}
\mbox{~~~~}
        \begin{tabular}{@{}l}
	#1
        \end{tabular}
	\\ \hline
}

\newenvironment{Pseudocode}{
      \begin{center}
      \begin{tabular}{ | p{0.95\textwidth} | } \hline
}
{
      \end{tabular}
      \end{center}
}

\newcommand{\PseudocodeAssertWStep}[2]{%
#1 & 
$%
\left\{ 
\colorbox{white}{
	\begin{minipage}{0.75\textwidth}
#2 
\end{minipage}%
}
\right\}  
$
\\ \hline 
}

\newcommand{\PseudocodeAssertIndentWStep}[2]{%
#1 &
\cellcolor{white}%
\mbox{~~~~} %
$%
\left\{ 
\cellcolor{white}{
	\begin{minipage}{0.73\textwidth}
#2
\end{minipage}
}
\right\}  
$
\\ \hline 
}

\newcommand{\PseudocodeStatementWStep}[2]{
\rowcolor{lightgray!25}
#1 & 
    \begin{tabular}{@{}l}
	#2
    \end{tabular}
	\\ \hline
}

\newcommand{\PseudocodeStatementIndentWStep}[2]{
\rowcolor{lightgray!25}
#1 &
\mbox{~~~~}
        \begin{tabular}{@{}l}
	#2
        \end{tabular}
	\\ \hline
}

\newenvironment{PseudocodeWStep}{
      \begin{center}
      \begin{tabular}{| c | p{0.8\textwidth} |} \hline
}
{
      \end{tabular}
      \end{center}
}

\newcommand{\FlaCostWorksheet}{
\begin{tabular}{| c | p{0.45\textwidth}
p{0.45\textwidth}|}\hline
Step & $\mbox{\color{blue}Algorithm:~}\routinename $ &
\\ \hline
1a &%
$ \left\{ 
\begin{minipage}{0.44\textwidth} 
$\ShowPrecondition$  
\end{minipage}
\right\}
$
&
\\ \hline
\rowcolor{lightgray!25}   
4 & %
\begin{minipage}{0.88\textwidth}%
\vspace{0.05in}
\ShowPartitionings~ \\
\mbox{\color{blue} ~~~where~} \ShowPartitionSizes
\end{minipage}
& 
\begin{minipage}{0.44\textwidth}
\hfill \CostInit
\end{minipage}
\\ \hline
2 & 
$ \left\{ 
\begin{minipage}{0.44\textwidth} 
$\ShowInvariant $
\end{minipage}
\right\} $ 
&
\begin{minipage}{0.44\textwidth}
$ \{ $ \hfill \CostInvariant
$ \} $ 
\end{minipage}
\\ \hline
\rowcolor{lightgray!25}   
3 &$\mbox{\color{blue}while~} \ShowGuard \mbox{~\color{blue} do}$
&
\\ \hline 
2,3 &  
$
\left\{
\begin{minipage}[t]{0.41\textwidth}%
$
~~~~ \ShowInvariant 
\wedge \ShowGuardTwo$
\end{minipage}
\right\}
$ 
&
\begin{minipage}{0.44\textwidth}
$ \{ $ \hfill \CostInvariant
$ \} $
\end{minipage}
\\ \hline
\rowcolor{lightgray!25}   
5a & ~~~~ \begin{minipage}{0.41\textwidth}%
\vspace{0.05in}
\ifthenelse{\equal{\blocksize}{1}}{}%
{%
\ifthenelse{ \equal{\blocksize}{blank} }{}%
{{\bf Determine block size $ \blocksize $}\\}%
}
\ShowRepartitionings~ 
\NoShow{\\
$\mbox{\color{blue} ~~~where~}$ \ShowRepartitionSizes}
\end{minipage}
&
\\ \hline
6 & 
$ \left\{ 
\begin{minipage}{0.41\textwidth} 
~~~~ \ShowBeforeUpdate 
\end{minipage}
\right\}
$
&
$ \{ $ \hfill 
\CostBefore
$ \} $
\\ \hline
\rowcolor{lightgray!25}  
8 & 
\begin{minipage}{0.41\textwidth} 
~~~~  \ShowUpdate 
\end{minipage}
&
\hfill 
$
C := C + 2 
$
\\ \hline 
7 & 
$ \left\{ 
\begin{minipage}{0.41\textwidth} 
~~~~ \ShowAfterUpdate 
\end{minipage}
\right\}
$
&
$ \{ $ \hfill 
\CostAfter
$ \} $
\\ \hline
\rowcolor{lightgray!25}   
5b & ~~~~ \begin{minipage}{0.41\textwidth}%
\vspace{0.05in}
\ShowMoveBoundaries~
\end{minipage}
&
\\ \hline
2 & 
$ \left\{ 
\begin{minipage}{0.44\textwidth} 
~~~~ $ \ShowInvariant  $ 
\end{minipage}
\right\}
$
&
\begin{minipage}{0.44\textwidth}
$ \{ $ \hfill \CostInvariant
$ \} $
\end{minipage}
\\ \hline
\rowcolor{lightgray!25}  
 &$\mbox{\color{blue} endwhile} $
 &
\\ \hline 
2,3 & 
$ \left\{ 
\begin{minipage}{0.44\textwidth} 
$ \ShowInvariant \wedge \neg( \ShowGuardTwo )$ 
\end{minipage}
\right\}
$
&
\begin{minipage}{0.44\textwidth}
$ \{ $ \hfill \CostInvariant
$ \} $
\end{minipage}
\\ \hline
1b & 
$ \left\{ 
\begin{minipage}{0.44\textwidth} 
$ \ShowPostcondition $ 
\end{minipage}
\right\}
$
&
\begin{minipage}{0.44\textwidth}
$ \{ $ \hfill \CostPostCond
$ \} $
\end{minipage}
\\ \hline
\end{tabular}
}

\newcommand{\leftdq}{\mbox{``}}
\newcommand{\rightdq}{\mbox{''}}
\newcommand{\PInv}{P_{\rm inv}}
\renewcommand{\wp}{\mbox{\rm wp}}

\usepackage{microtype}

\title{A Simple Methodology for Computing Families of Algorithms \\[0.2in]
\large
FLAME Working Note \#87}

\NoShow{
\author{Margaret E. Myers}{Department of Statistics and Data Sciences, The University of Texas at Austin}{myers@cs.utexas.edu}{}{}

\author{Richard Vuduc}{School of Computational Science and Engineering, Georgia Institute of Technology}{richie@cc.gatech.edu}{}{}

\author{Robert A. van de Geijn}{Institute for Computational Engineering and Sciences, The University of Texas at Austin}{rvdg@cs.utexas.edu}{}{}

\authorrunning{D. N. Parikh, M. E. Myers, R. Vuduc, and R. A. van de Geijn}

\Copyright{Devangi N. Parikh, Margaret E. Myers, Richard Vuduc, and Robert A. van de Geijn}

\subjclass{\ccsdesc[500]{Theory of computation~Algorithm design techniques}}

\keywords{program correctness, formal derivation, program synthesis, cost analysis, polynomial evaluation,
loop-based algorithms}

\category{}

\relatedversion{}

\supplement{}

\funding{This research was supported in part by NSF under grant numbers ACI-1550493, and ACI-1714091.}

\acknowledgements{We would like to thank our many collaborators who have contributed to this research over the years.}

\EventLongTitle{2nd Symposium on Simplicity in Algorithms (SOSA 2019)}
\EventShortTitle{SOSA 2019}
\EventAcronym{SOSA}
\EventYear{2019}
\EventDate{January 6--9, 2019}
\EventLocation{San Diego, California}
\EventLogo{}
}

\newcommand*{\affaddr}[1]{#1} 
\newcommand*{\affmark}[1][*]{\textsuperscript{#1}}
\newcommand*{\email}[1]{\texttt{#1}}

\begin{document}

\author{
	Devangi N. Parikh\affmark[a],
	Margaret E. Myers\affmark[a],
	Richard Vuduc\affmark[b]\footnote{Supported in part by the J. Tinsley Oden Faculty Fellowship.},
	Robert A. van de Geijn\affmark[a]\footnote{Supported in part by the 2016 Peter O'Donnell Distinguished Researcher Award.}\\
	\affaddr{\affmark[a]The University of Texas at Austin, Austin, TX }\\
	\affaddr{\affmark[b]Georgia Institute of Technology, Atlanta, GA}\\
	\email{{\tt \{dnp@cs.utexas,  myers@cs.utexas, richie@cc.gatech.edu, rvdg@cs.utexas.edu\}}}
}

\date{August 20, 2018}

\maketitle

\begin{abstract}
    Discovering ``good'' algorithms for an operation is often considered an art best left to experts. 
What if there is a simple methodology, an algorithm, for systematically deriving a family of algorithms as well as their cost analyses, so that the best algorithm can be chosen?
We discuss such an approach for deriving loop-based algorithms.  The example used to illustrate this methodology, evaluation of a polynomial, is itself simple yet the best algorithm that results is surprising to a non-expert: Horner's rule. 
We finish by discussing recent advances that make this approach highly practical for the domain of high-performance linear algebra software libraries.

\end{abstract}

\section{Introduction}
Finding a ``good'' algorithm by some measure (e.g., time complexity, wall-clock time, or simplicity) often means picking  from a family of known algorithms. 
The question becomes how to find the members of such a family of algorithms, given the specification of the operation to be computed.
Often this is viewed as an art.  If instead the process can be made systematic, it becomes a science.
Determining algorithms for a specified problem is itself a computation, one that takes a specification as input and yields algorithms for computing it as output.
Thus, the goal is to find an
algorithm,
which we  call the 
FLAME
methodology~\cite{FLAME}, for computing families of algorithms.

\NoShow{What we ideally want is a simple algorithm for finding algorithms that can be taught and implemented.  Thus art becomes science.}

\subsection{A worksheet for deriving loop-based algorithms}

Our own research focuses on high-performance computing.  A ``good'' algorithm in that area uses the memory hierarchy efficiently.  What this often means is that algorithms are loop-based, where the stride
(block size) through the data (vectors and matrices) are chosen so as to nearly optimally amortize the cost of moving data between memory layers.
Hence, we focus this paper on a methodology for deriving  families of loop-based algorithms.

The approach starts with the idea of programming loops as
what David Gries calls a goal-oriented activity -- built on the foundations of Dijkstra, Hoare, and others~\cite{Hoare:1969:ABC:363235.363259,Dijkstra:1975:GCN:360933.360975,Dijkstra,Floyd:67}, and elegantly synthesized by Gries~\cite{Gries} -- wherein proofs of correctness and programs are developed together. However, the process is made more systematic by organizing the approach in what we call a ``worksheet'' that clearly links the derivation to the assertions that must hold at different points in the algorithm.

\NoShow{\bf Complete this paragraph, ending with  Horner's rule here.  I suggest we don't actually explain the ``simple'' explanation of Horner's rule until after Section 4.  Should we show the algorithm here without explanation, much like we do in the MOOC?}

\subsection{Related work} 

There is a long history of related work in the human-guided synthesis of programs from proofs, i.e., deductive synthesis systems backed by theorem provers~\cite{Manna:1980ds,Constable:1986nu,Smith:1990ki}. These are quite powerful and sophisticated. As noted by others~\cite{Solar-Lezama:2008sk}, they also require the human programmer to have a deep mathematical background, in order to formulate theorems and nudge the theorem prover along. In an effort to make formal synthesis systems more usable, several semi-automatic systems have taken the approach of ``changing the interface'' between the programmer and proof system from derivations to partial programs~\cite{Thies:2002dp,Fischer:2003ab,co-induction-simply-automatic-co-inductive-proofs-in-a-program-verifier}, which includes a state-of-the-art technique known as sketching~\cite{Solar-Lezama:2009sk}. However, an end-user programmer of these systems must still intuit the partial program, and the synthesis process may require unit testing against at least one reference implementation. By contrast, our methodology tries to strike a balance between the approaches of deductive synthesis and partial programs. In particular, our worksheet formalism structures both program \emph{and} proof, with the methodology (meta-algorithm) providing the steps for the programmer to follow.

\subsection{An illustrating example}

The simple example that we use to illustrate the ideas is the evaluation of a polynomial given its coefficients and a value:
\[
y := a_0 + a_1 x + \cdots + a_{n-1} x^{n-1},
\quad
\mbox{or, equivalently,}
\quad
y := \sum_{i=0}^{n-1} a_i x^i .
\]
While the described methodology may seem like overkill for operations over vectors of data (the coefficients in this example), we find that polynomial evaluation is already quite interesting, 
yielding a novel derivation of the non-trivial Horner's Rule%
\footnote{While named after William G. Horner, this algorithm was actually discovered earlier, as discussed in the Wikipedia article on the subject.}%
~\cite{Horner:1819pe} in a form accessible to a novice. \NoShow{Indeed, several of this paper's authors have even used this worksheet in a number of courses, with undergraduates unknowingly deriving a large fraction of the functionality of available in a modern dense linear algebra library (Section~\ref{sec:enrichments})."}

\subsection{Contributions}

This paper makes a number of contributions:
\begin{itemize}
    \item 
    It revisits the importance of formal derivation of loop-based algorithms and raises awareness of advances regarding its practical importance.
    \item
    It structures such derivations as a simple algorithm.
    \item
    It demonstrates the power of abstractions that hide intricate indexing.
    \item
    It illustrates the insights using a simple example appropriate for beginning computer science students.
    \item
    While the example appears in the Massive Open Online Course (MOOC) titled ``LAFF-On Programming for Correctness,'' with explicit indexing into arrays~\cite{LAFF-On,LAFF-On-edX}, the derivation using the FLAME notation in Section~4 is new.
    \item
    It introduces a systematic approach to deriving the cost analysis of a loop-based algorithm.
    \end{itemize}
Along the way, we encounter simplicity at multiple levels.

\section{Deriving algorithms using classical notation}

~ \hfill
\begin{minipage}[c]{4.5in}
The only effective way to raise the confidence
level of a program significantly is to give a convincing proof of its correctness. But one should
not first make the program and then prove its
correctness, because then the requirement of
providing the proof would only increase the poor
programmers burden. On the contrary: {\color{red} the programmer should let correctness proof and program grow hand in hand.}

~ \hfill
Dijkstra (1972) ``The Humble Programmer'' 
\end{minipage}
\vspace{0.2in}

\subsection{Basic tools---Review of concepts}

The following three definitions are foundational to deriving algorithms.

\begin{description}
\item[The Hoare Triple:] Given predicates $Q$ and $R$, the Hoare triple $\{Q\} S \{R\}$ holds if and only if
executing the command $ S $ starting in a state that satisfies $ Q $ completes in a finite time in a state that satisfies $ R $. Here, $Q$ and $R$ are known as the precondition and postcondition, respectively~\cite{Hoare:1969:ABC:363235.363259,Gries}. 

\item[The Weakest Precondition:] Given command $S$ and postcondition $R$, the weakest precondition, denoted by $\wp(``S", R)$, is the weakest predicate, $ Q $, such that 
$ \{ Q \} S \{ R \} $ holds~\cite{Dijkstra:1975:GCN:360933.360975,Dijkstra,Gries}.
In other words, it describes the set of all states of the variables for which execution of $S$ in a finite time results in a state for which postcondition $R$ holds. The definition of weakest precondition allows us to prove that a segment of code is correct: $ \{ Q \} S \{ R \} $ if and only if $ Q \Rightarrow \wp( ``S", R ) $.

\item[The Loop Invariant:] A loop invariant, $\PInv$, is a predicate that is true at the beginning as well as at the end of each iteration of the loop~\cite{Floyd:67,Gries}.
In our discussion, the loop invariant describes {\em all} pertinent variables needed to prove the loop correct.
\end{description}

\subsection{A worksheet for proving loops correct}

A generic while loop is given by

\vspace{0.2in}
\begin{minipage}[h]{\textwidth}
	{\bf while} $G$ {\bf do} \\
	 \mbox{~~} $S$ \\
 	{\bf endwhile} 
\end{minipage}
\vspace{0.2in}

\newcommand{\while}{\mbox{\rm WHILE}}

\noindent We will denote this while loop by \while.  Given a precondition, $ P_{\rm pre} $, and postcondition, $ P_{\rm post} $, we would like to derive $ S$ and $ G $ so that $ \{ P_{\rm pre} \} \while \{ P_{\rm post} \} $ is true.

\renewcommand{\precondition}{P_{\rm pre}}
\renewcommand{\postcondition}{P_{\rm post}}
\renewcommand{\invariant}{\PInv}

\newcommand{\loopguard}{{G}}
\begin{figure}[tb!]
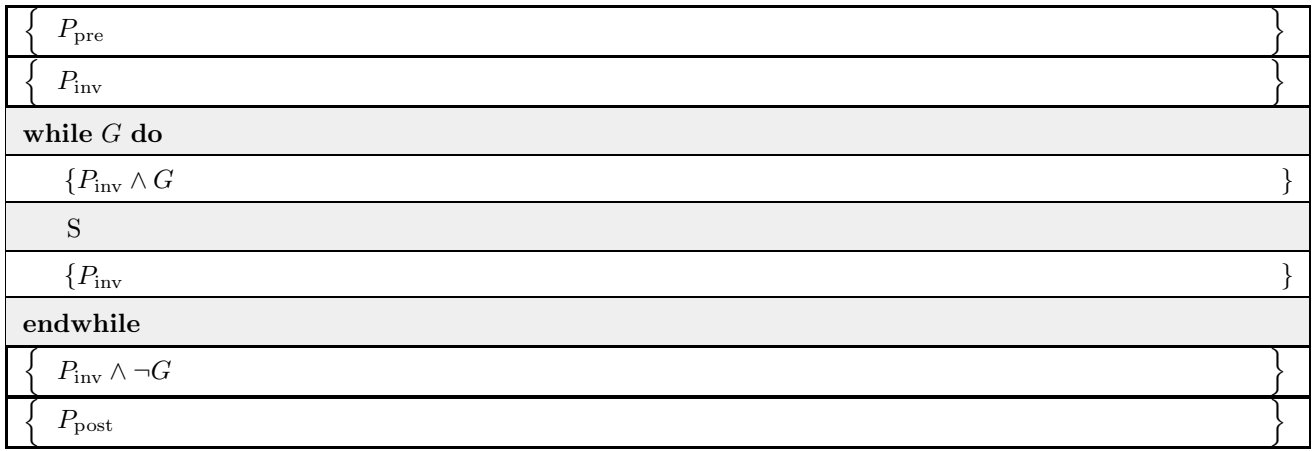

	\begin{Pseudocode}
		\PseudocodeAssert{ $P_{\rm pre} $}
		\PseudocodeAssert{$ \PInv$}
		\PseudocodeStatement{ {\bf while} $ \loopguard $  {\bf do}} 
		\PseudocodeAssertIndent{$ \PInv \wedge \loopguard$}
		\PseudocodeStatementIndent{S}
		\PseudocodeAssertIndent{$\PInv $}
		\PseudocodeStatement{ {\bf endwhile} } 
		\PseudocodeAssert{$\PInv \wedge \neg \loopguard$}
		\PseudocodeAssert{$ P_{\rm post} $}
	\end{Pseudocode}
	\caption{A generic while loop annotated with assertions that facilitates the proof of correctness of the loop.}
	\label{fig:annotatedLoop}
\end{figure}

\noindent We start by discussing how to prove that WHILE is correct.
Figure~\ref{fig:annotatedLoop} shows the loop annotated with assertions (in the grey boxes) that can be used to prove the loop correct, using the Principle of Mathematical Induction (PMI).
For those familiar with how loops can be proved correct, this annotated algorithm captures the DO theorem%
\footnote{For now, we will deal with what is known as partial correctness: we will not prove that the loop completes.}~\cite{Gries}:
\begin{itemize}
    \item 
    If $ P_{\rm pre} \Rightarrow \PInv $, then we know that $ \PInv $ holds before the loop starts. This is the base case for our inductive proof.
    \item
    The inductive step requires proving that $ \{ \PInv \wedge G \} S \{ \PInv \} $.  What this says is that {\em if} before $ S $ is executed the variables are in a state that satisfies $ \PInv $ {\em and} the loop guard $ G $ then the variables satisfy $ \PInv $ after the execution of $ S $.  
    \item
    By the PMI, we then know that $ \PInv $ is true before and after the execution of $ S $ as long as the loop continues to execute ($ G $ evaluates to {\em true}).
    \item
    {\em If} the loop ever completes, then $ \PInv $ is still true after the loop {\em and} $ \neg G $ must be {\em true}.  
    \item
    If $ \PInv \wedge \neg G \Rightarrow P_{\rm post} $, then we finish this code segment in a state where $ P_{\rm post} $ is {\em true}.  
    \item
    We conclude that {\em if} the loop finishes, then $ \{ P_{\rm pre} \} S \{ P_{\rm post}\} $ holds.  In other words, the code segment is correct.
    \end{itemize}
\NoShow{
if the loop invariant 
$\PInv$ holds before the {\bf while} loop begins, at every iteration of the loop, and once the loop has terminated then the loop is said to be correct.\footnote{Or, partially correct to be exact. Further proof is required to prove the loop terminates.} This follows by applying the Principle of Mathematical Induction.
}
This discussion assumes that we know $ P_{\rm pre} $, $ \PInv $, $ G $, $ S $, and $ P_{\rm post} $.
What we want instead is a methodology (algorithm) for deriving (computing)  $ \PInv $, $ G $, and $ S $,
given $ P_{\rm pre} $ and $ P_{\rm post} $.

\subsection{A worksheet for deriving correct loops}

We now transform this annotated {\bf while} loop for proving the loop's correctness into a worksheet to be filled out
(an algorithm to be executed) in order to \emph{derive} a correct loop, guided by its proof of correctness. 
It is an algorithm for goal-oriented programming.

\begin{figure}[tb!]
    \begin{PseudocodeWStep}
        Step & Operation \\ \hline
		\PseudocodeAssertWStep{1a}{ $\precondition$}
		\PseudocodeAssertWStep{4a}{ $\wp( \leftdq S_I \rightdq, \PInv ):$}
		\PseudocodeStatementWStep{4b}{$S_I$}
		\PseudocodeAssertWStep{2}{$ \PInv$}
		\PseudocodeStatementWStep{3}{ {\bf while} $ \loopguard $  {\bf do}} 
		\PseudocodeAssertIndentWStep{2,3}{$ \PInv \wedge G$}
		\PseudocodeAssertIndentWStep{7}{$\wp \left(\leftdq S_U; k:=k-1 \rightdq,\PInv \right)$}
		
		\PseudocodeStatementIndentWStep{8}{$S_U$}
		\PseudocodeAssertIndentWStep{6}{$\wp \left(\leftdq k:=k-1 \rightdq,\PInv \right)$}
		\PseudocodeStatementIndentWStep{5}{$k:= {\cal E}_{k} $}
		\PseudocodeAssertIndentWStep{2}{$\PInv$}
		\PseudocodeStatementWStep{}{ {\bf endwhile} } 
		\PseudocodeAssertWStep{2,3}{$\PInv \wedge \neg G$}
		\PseudocodeAssertWStep{1b}{$\postcondition$}    
    \end{PseudocodeWStep}
    \caption{Template of a blank worksheet used to derive algorithms to be correct.}
    \label{fig:blankWorksheetIndx}
\end{figure}

Given the precondition and postcondition, if the loop invariant $\PInv$ is derived from these, then the rest of the loop can be determined systematically. 
The order in which the derivation should proceed is indicated by the step numbers in Figure~\ref{fig:blankWorksheetIndx}.  Thus, the annotated \while\ command in Figure~\ref{fig:blankWorksheetIndx} becomes a worksheet to be filled out in order to derive the loop.

There are a number of assumptions embedded in this worksheet: 
\begin{itemize}
    \item
    There is an initialization step $ S_I $.
    \item
    The loop index is $ k $. It is updated at the bottom of the loop body so that the loop body $ S $ becomes $ S_U ; k := {\cal E}_k $ (update command $ S_U $ followed by the update of $ k $ with expression $ {\cal E}_k $).
    \end{itemize}
Similar worksheets can be created for other prototypical loops, e.g., one that updates a loop index at the top of the loop, as discussed in the MOOC~\cite{LAFF-On,LAFF-On-edX}.

We next demonstrate the methodology with a simple example.

\subsection{Algorithms for Evaluating Polynomials}
\label{subsec:evalPolyIdx}
By systematically filling out the worksheet one can discover a family of algorithms to evaluate a polynomial.
This family includes the well-known Horner's Rule~\cite{Horner:1819pe}, which is optimal in the sense of performing the least number of additions and multiplications\footnote{The optimality analysis assumes no ``preconditioning'' of coefficients~\cite{Pan:1966py}.}~\cite{Pan:1966py,Knuth:1998a-vol2}.
However, discovering Horner's Rule is not obvious, so it may be surprising that the systematic procedure illustrated in this section can in fact uncover it.
The steps follow the numbering of the left column in Figure~\ref{fig:blankWorksheetIndx}.

\NoShow{\renewcommand{\precondition}{0 \leq n}
\renewcommand{\postcondition}{y = \left(\sum i \, | \, 0 \leq k < n : y = a_i x^{i}\right) }
\renewcommand{\invariant}{y = \left(\sum i \, | \, k \leq i < n \, : \,  a_i x^{i-k}  \right) \wedge  \, \left(0 \leq k \leq n\right)}
\renewcommand{\loopguard}{{\left(k > 0\right)}}
}

\renewcommand{\precondition}{0 \leq n}
\renewcommand{\postcondition}{y = \sum_{i=0}^{n-1} a_i x^i }
\renewcommand{\invariant}{y = \sum_{i=k}^{n-1} a_i x^{i-k} \wedge \, 0 \leq k \leq n }
\renewcommand{\loopguard}{{\left(k > 0\right)}}

\subsubsection{Step 1: Precondition and postcondition.} To derive an algorithm we must first specify the precondition and the postcondition.
To evaluate a polynomial  $y = a_0 + a_1 x + a_2 x^2 ... +a_{n-1} x^{n-1}$,  the precondition can be given by 
\[P_{\rm pre}: \precondition,\] and the postcondition by \[P_{\rm post}: \postcondition.\] 
We (implicitly) assume that the coefficients $ a_i $ are stored in array $ a $ 
of appropriate size, $ n $.

\subsubsection{Step 2: Determine loop invariants.} 
The question is now is how to systematically determine loop invariants for a given operation. By its very nature, the loop invariant should capture partial progress towards the postcondition.
Splitting the range of the summation in the postcondition yields 
 \begin{equation*}
y =  \sum_{i=0}^{k-1} a_i x^i  + \sum_{i=k}^{n-1} a_i x^i ,\,0 \leq k \leq n 
\end{equation*} 
and with some algebraic manipulation,  
\begin{equation}
\label{eqn:pmi}
y = \sum_{i=0}^{k-1} a_i x^i + \left(\sum_{i=k}^{n-1} a_i x^{i-k} \right)x^k  \wedge \, 0 \leq k \leq n.
\end{equation}  
Why do we split the range?  Many algorithms over arrays traverse those arrays inherently from the first element to the last or from the last element to the first.  The splitting captures that the computation has reached a point where $ k $ elements have been so processed (if the algorithm traverses from first to last) or are left to be processed (if traversed from last to first).
We recognize (\ref{eqn:pmi}) as a recursive definition of the operation. 

\begin{figure}[tb!]
\begin{center}
\begin{tabular}{|l I l l |} \hline
	Invariant 1 & $y = \sum_{i=0}^{k} a_i x^i $&$ \wedge  \, 0 \leq k \leq n$\\\hline
	Invariant 2 & $y = \sum_{i=k}^{n-1} a_i x^i \, 
	$&$
	\wedge  \, 0 \leq k \leq n$ \\\hline
	Invariant 3 & $y = \sum_{i=0}^{k} a_i x^i\,\wedge z= x^k\, $&$
	\wedge  \, 0 \leq k \leq n$\\ \hline
	Invariant 4 & $y = \sum_{i=k}^{n-1} a_i x^i\,\wedge z= x^k\, $&$
	\wedge  \, 0 \leq k \leq n $\\ \hline
	Invariant 5 & $y = \sum_{i=k}^{n-1} a_i x^{i-k} $&$
	\wedge  \, 0 \leq k \leq n $\\ \hline
\end{tabular}
\vspace{0.1in}
\caption{Loop invariants for evaluating a polynomial}
\label{fig:loopInvIndx}
\end{center}
\end{figure}

When control is at the top or bottom of the loop, a partial result toward the final result has been computed.
Many subexpressions of (\ref{eqn:pmi}) represent partial progress towards the postcondition.  
That guides us towards a systematic way of identifying possible loop invariants: pick subexpressions of (\ref{eqn:pmi}).%
\footnote{Some subexpressions do not yield valid loop invariants. For example, $ y = 0 \wedge 0 \leq k \leq n-1 $ is a subexpression that yields a loop that does not compute anything.  How to deal with this goes beyond this paper as does the question whether Figure~\ref{fig:loopInvIndx} gives all loop invariants.} This yields at least the five different loop invariants listed in Figure~\ref{fig:loopInvIndx}. 

Suppose we choose Invariant 5.
Then $\PInv: \invariant$ 
is entered everywhere it should hold in the worksheet of Figure~\ref{fig:blankWorksheetIndx}, i.e., in the four places marked with Step 2.

\subsubsection{Step 3: Determine the loop guard.} 
From the bottom of the worksheet, we have:
\begin{PseudocodeWStep}
	\PseudocodeAssertWStep{2,3}{\shortstack[l]{$\PInv \wedge \neg G:$  $ \invariant \wedge \neg G$}}
	\PseudocodeAssertWStep{1b}{$ P_{\rm post}: \postcondition$}
\end{PseudocodeWStep}
For this to be correct, $\PInv \wedge \neg G$ must imply $\postcondition$. This dictates that $G$ must be chosen as $k > 0$.
This is then entered for $ G $ in the worksheet, everywhere Step~3 appears.

\subsubsection{Step 4: Initializing the loop.} 

Informally, starting in a state described by the precondition $\precondition$, to place the variables in a state where the loop invariant $\invariant$ is true, $k$ and $y$ must be initialized to $n$ and $0$, respectively.

More formally, 
we start by choosing $ S_I: y,k := {\cal E}_0, {\cal E}_1 $ (simultaneous assignment of expressions $ {\cal E}_0 $ and $ {\cal E}_1 $ to $ y $ and $ k $), where $ {\cal E}_0 $ and $ {\cal E}_1 $ are expressions that are to be determined.
Then
\begin{eqnarray*}
\nonumber
\wp(``S_I", \PInv) &\mbox{is}&
\wp( ``y,k := {\cal E}_0, {\cal E}_1", \invariant )\\
\nonumber
&\mbox{is}&
\mbox{$ {\cal E}_0
= \sum_{i={\cal E}_1}^{n-1} a_i x^{i-{\cal E}_1} \wedge 0 \leq {\cal E}_1 \leq n$}
\end{eqnarray*}
by the definition of the weakest precondition of simultaneous assigment.
Since $ \precondition$ must imply 
$ \wp(``S_I", \PInv) $
we deduce that $ y, k := 0, n $.
We can recognize that we can 
instead make two separate assignments: 
\[ S_I: y := 0 ; k := n .\]
This is entered as Step~4.

\subsubsection{Step 5: Progressing through the loop.} To make progress towards completing the computation, $ k $ must eventually equal $ 0 $ so that the loop guard becomes false. Since we start the loop at $k:=n$, this means $k$ must be decremented in each iteration of the loop.  This leads us to choose $ {\cal E}_k $ to equal $ k -1 $.
This is entered as Step~5.

\subsubsection{Step 6: Weakest precondition of the indexing statement.} 

What is left is to determine $ S_U $.  According to the worksheet,
\begin{PseudocodeWStep}
	\PseudocodeAssertWStep{6}{$\wp\left(\leftdq k:= k-1 \rightdq, \PInv\right)$}
	\PseudocodeStatementWStep{5}{$k:=k-1$}
	\PseudocodeAssertWStep{2}{$\PInv: \invariant $}
\end{PseudocodeWStep}
For this code segment to be correct, the weakest precondition $\wp\left(\leftdq k:= k-1 \rightdq, \PInv\right)$ must hold at Step~6. Simplifying, we get
\begin{align*}
& \wp\left(\leftdq k:= k-1 \rightdq, \PInv\right) \\
\Leftrightarrow & < \textrm{\textit {Instantiate}}\, \PInv > \\
& \wp\left(\leftdq k:= k-1 \rightdq, \invariant \right) \\
\Leftrightarrow & < \textrm{\textit {Definition of}}\, {\leftdq}:={\rightdq} > \\
& y = \sum_{i=k-1}^{n-1} a_i x^{i-(k-1)} \wedge \, 0 \leq k-1 \leq n \\
\Leftrightarrow & < \textrm{\textit {Splitting the range and algebra}}> \\
& y = a_{k-1} + \left( \sum_{i=k}^{n-1} a_i x^{i-k} \right) x\wedge \, 1 \leq k \leq n+1.
\end{align*}
This is entered as Step~6.

\subsubsection{Step 7: Weakest precondition of the update statement.} 

Working our way up the loop body further, we now notice that $ y $ must be updated by some expression: 
\begin{PseudocodeWStep}
    \PseudocodeAssertWStep{7}{$\wp\left( \leftdq y:={\cal E} \rightdq ,y = a_{k-1} + \left( \sum_{i=k}^{n-1} a_i x^{i-k} \right) x\wedge \, 1 \leq k \leq n+1\right)$}
    \PseudocodeStatementWStep{8}{$y:={\cal E}$}
	\PseudocodeAssertWStep{6}{$y = a_{k-1} + \left( \sum_{i=k}^{n-1} a_i x^{i-k} \right) x\wedge \, 1 \leq k \leq n+1$}
\end{PseudocodeWStep} 
We do not know the update statement yet, but we know it will be of the form $y := {\cal E}$.
%
From this fact, we can derive the weakest precondition for this segment of the loop.
\begin{align*}
& \wp\left( \leftdq y:= {\cal E} \rightdq ,y = a_{k-1} + \left( \sum_{i=k}^{n-1} a_i x^{i-k} \right) x\wedge \, 1 \leq k \leq n+1\right) \\
\Leftrightarrow & < \textrm{\textit {Definition of}}\, {\leftdq}:={\rightdq} > \\
& {\cal E} = a_{k-1} + \left( \sum_{i=k}^{n-1} a_i x^{i-k} \right) x\wedge \, 1 \leq k \leq n+1
\end{align*}

{\bf Step 8: Loop update.} Lastly, to determine the update statement, consider the topmost part of the loop:
\begin{PseudocodeWStep}
	\PseudocodeAssertWStep{2,3}{$ \PInv \wedge G: \invariant \wedge \loopguard$}
	\PseudocodeAssertWStep{7}{${\cal E} = a_{k-1} + \left( \sum_{i=k}^{n-1} a_i x^{i-k} \right) x\wedge \, 1 \leq k \leq n+1$}
\end{PseudocodeWStep}
Since, $ \PInv \wedge G$ must be stronger than Step 7, by comparison of terms, we deduce that
\[
{\cal E} = a_{k-1} + y \times x
\]
since
\[
\begin{array}{l}
( \invariant \wedge ( k > 0 ) 
\Rightarrow \\
~~~~( a_{k-1} + y \times x = a_{k-1} + \left( \sum_{i=k}^{n-1} a_i x^{i-k} \right) x\wedge \, 1 \leq k \leq n+1.
\end{array}
\]
so that $ S_U $ becomes $ y := a_{k-1} + y \times x $, which is filled in for Step~8.

The complete algorithm and its proof of correctness are summarized in Figure~\ref{fig:HornerIndx}.

\begin{figure}[bt!]
	\begin{PseudocodeWStep}
	    Step & Operation \\ \hline
		\PseudocodeAssertWStep{1a}{ $ P_{\rm pre}: \precondition$}
		\PseudocodeAssertWStep{4a}{ $\wp( \leftdq S_I \rightdq, \PInv ):
		{\cal E}_0
= \sum_{i={\cal E_1}}^{n-1} a_i x^{i-{\cal E}_1} \wedge 0 \leq {\cal E}_1 \leq n $}
		\PseudocodeStatementWStep{4b}{\shortstack[l]{$S_I: k:=n;y:= 0$}}
		\PseudocodeAssertWStep{2}{$ \PInv: \invariant$}
		\PseudocodeStatementWStep{3}{ {\bf while} $ \loopguard $  {\bf do}} 
		\PseudocodeAssertIndentWStep{2,3}{\shortstack[l]{$ \PInv \wedge G: \invariant \wedge \loopguard$}}
		\PseudocodeAssertIndentWStep{7}{\shortstack[l]{$\wp \left(\leftdq y= {\cal E}; k:={\cal E}_k \rightdq,\PInv \right):$ \\
		$ ~~~~~{\cal E} = a_{k-1} + \left(\sum_{i=k}^{n-1} a_i x^{i-k}\right) x \wedge \, 1 \leq k \leq n+1$}}
		
		\PseudocodeStatementIndentWStep{8}{$ S_U: y := \begin{array}[t]{c}
		\underbrace{a_{k-1} + y \times x}\\ {\cal E}
		\end{array}$}
		\PseudocodeAssertIndentWStep{6}{\shortstack[l]{$\wp \left(\leftdq k:=
		{\cal E}_k
		\rightdq,\PInv \right): y = a_{k-1} + \left(\sum_{i=k}^{n-1} a_i x^{i-k}\right) x \wedge \, 1 \leq k \leq n+1$}}
		\PseudocodeStatementIndentWStep{5}{$k:=\begin{array}[t]{c}
		\underbrace{k-1}\\
		{\cal E}_k 
		\end{array}$}
		\PseudocodeAssertIndentWStep{2}{\shortstack[l]{$\PInv: \invariant $}}
		\PseudocodeStatementWStep{}{ {\bf endwhile} } 
		\PseudocodeAssertWStep{2,3}{\shortstack[l]{$\PInv \wedge \neg G: \invariant \wedge \neg \loopguard$}}
		\PseudocodeAssertWStep{1b}{$P_{\rm post}: \postcondition$}
	\end{PseudocodeWStep}
	\caption{Worksheet used to systematically derive Horner's Rule.}
    \label{fig:HornerIndx}
\end{figure}

\NoShow{
\begin{align*}
  &\wp \left( ``k:=k-1; y:= {\cal E}'', \PInv \right) \\
= &\wp \left( ``k:=k-1; y:= {\cal E}'', \invariant \right) \\
= &\wp \left( ``k:=k-1'', {\cal E} = \left(\sum i \, | \, k \leq i < n \, : \,  a_i x^{i-k}  \right) \wedge  \, \left(0 \leq k \leq n\right) \right) \\
= & E(k-1) = \left(\sum i \, | \, k-1 \leq i < n \, : \,  a_i x^{i-\left( k-1 \right)}  \right) \wedge  \, \left(0 \leq k-1 \leq n\right)
\end{align*}
}

\subsection{Discussion}
The procedure in the previous section can be repeated for the other loop invariants listed in Table~\ref{fig:loopInvIndx}, yielding a family of algorithms. It is left to the interested reader to derive the remaining algorithms.

The worksheet is not unique. Placing the update of the loop index at the top of the loop  results in yet more algorithms, albeit via a similarly systematic procedure.

\subsection{Alternative explanation of Horner's rule}

Now that we have derived the algorithm, we can recognize that
\[
a_0 + a_1 x + \cdots + a_{n-1} x^{n-1}
=
a_0 + ( a_1 + ( a_2 + ( \cdots (a_{n-1}+0 \times x ) \times x \cdots ) \times x ) \times x , 
\]
which also explains Horner's Rule.
  {\bf The methodology yielded the algorithm that captures this insight without first having the insight.}

\section{Deriving loops using the FLAME notation}
\label{sec:FLAME}
~ \hfill
\begin{minipage}[c]{4.5in}
How do we convince people that in programming simplicity and clarity — in short: what mathematicians call ``elegance'' — are not a dispensable luxury, but a crucial matter that decides between success and failure?

~ \hfill
Dijkstra (1982)~\cite{EWD648} 
\end{minipage}
\vspace{0.2in}

Deriving algorithms as described in the previous section, using a notation in which indices are explicit both in the code and in the quantifiers that appear in the assertions, may seem cumbersome and counterintuitive. 
Moreover, the opportunity for introducing errors in the derivation because of indexed expressions is akin to the opportunity for introducing errors in indexed loops in the first place.
In this section, we will show that using a compact notation that hides indices makes deriving
and understanding algorithms easier.

\subsection{Basic tools---review of notation}

\renewcommand{\operation}{\psi = \pi( a, \chi )}

\renewcommand{\routinename}{\operation}

\renewcommand{\precondition}{{0 \leq n}}

\renewcommand{\postcondition}{{\psi = pi(a, \chi)}}

\renewcommand{\invariant}{{\psi = p\left(a_B, \chi \right)}}
\renewcommand{\blocksize}{blank}

\renewcommand{\guard}{{m(a_B) < m(a)}}

\renewcommand{\partitionings}{{$
\begin{array}{@{}l}
\psi:=0 \\ a \rightarrow \FlaTwoByOne{a_T}{a_B}
\end{array}
$}}

\renewcommand{\partitionsizes}{$  a_B  $ has  0  elements}

\renewcommand{\repartitionings}{
$
{\FlaTwoByOne{a_T}{a_B}
	\rightarrow
	\FlaThreeByOneT{a_0}{\alpha_1}{a_2}}   
$
}

\renewcommand{\repartitionsizes}{$ a_B $ has  0  elements}

\renewcommand{\moveboundaries}{
$
{\FlaTwoByOne{a_T}{a_B}
	\leftarrow
	\FlaThreeByOneB{a_0}{\alpha_1}{a_2}}    
$
}

\renewcommand{\beforeupdate}{{$\psi = \pi \left(a_2, \chi \right)$}}

\renewcommand{\afterupdate}{{$\psi = \pi \left(\alpha_1, \chi \right) + \pi \left(a_2, \chi \right) \chi$}}

\renewcommand{\update}{
{$
\begin{array}{l}
    \psi := \alpha_1 + \psi \times \chi
\end{array}
$}
}

\begin{figure}[tb!]
\begin{center}
\begin{tabular}{c c}
\FlaAlgorithm
&
\begin{minipage}{3.25in}
\begin{itemize}
    \item 
    For added clarity, we 
    use lower case Roman letters for vectors and lower case Greek letters for scalars.
    Thus, what were $ x $ and $ y $ become $ \chi $ (chi) and $ \psi $ (psi), and the element of vector $ a $ are denoted with $ \alpha $.
    \item
    The function $ \pi( a, \chi ) $ computes the polynomial with coefficients given in vector $ a $ evaluated at $ \chi $.  We are to compute $ \psi = \pi( a, \chi )$.
    \item
    The algorithm starts by initializing $ \psi $ to zero and partitioning the vector into a subvector of elements with which we will compute in the future, $ a_T $, and a subvector of elements with which we have already computed, $ a_B $.  Initially, the second vector has no elements.
    \item
    The function $ m( a ) $ returns the size of vector $ a $.
    \item
    The body of the loop first exposes the last element of $ a_T $. It then updates $ \psi $ with that exposed element.  Finally, it adds the exposed element to the subvector of coefficients with which computation has completed.
    \end{itemize}
\end{minipage}
\end{tabular}
\end{center}
\caption{Horner's Rule for computing $ \psi = \pi( a, x ) $ in FLAME notation.}
\label{fig:HornersRuleFLAME}
\end{figure}

We now introduce an alternative notation, referred to as the FLAME notation~\cite{FLAME,TSoPMC,CiSE09}, that we have been using in our research, development, and educational outreach~\cite{LAFF,LAFF-edX,LAFF-On,LAFF-On-edX}, for almost two decades.
To do so, we present the algorithm for Horner's Rule 
with this notation, in Figure~\ref{fig:HornersRuleFLAME}.
The notation hides the details of indexing.
It should be intuitively obvious how, for example, an algorithm that marches through vector $ a $ from top to bottom would be expressed.

\NoShow{
The coefficients of the polynomial are stored in a vector, $ a $.
In our discussion of evaluating a polynomial, the coefficients $ a_0, a_1, \cdots $ are elements in a vector of data, $ a $.
To distinguish variables that are vectors (1D arrays, viewed as a column of elements), going forward we will use lower case Greek letters for scalars and lower case Roman letters for vectors%
\footnote{Matrices (2D arrays) are denoted by upper case Roman letters.  In this paper, we will not discuss operations with matrices.}.

We noticed in our discussion of polynomial evaluation that the vector of coefficients
\[
a = \left( \begin{array}{c}
\alpha_0 \\
\alpha_1 \\
\vdots \\
\alpha_{n-1}
\end{array}
\right) ,
\]
(now using lower case Greek letters for the scalar coefficients)
can be viewed as the coefficients with which we have already computed, 
$ \alpha_{k}, \ldots , \alpha_{n-1} $ 
and those with which we have yet to compute, 
$ \alpha_{0}, \ldots , \alpha_{k-1} $.
In order to hide the indices needed to express these ranges, we instead view $ a $ as being partitioned into subvectors
\[
a = \left( \begin{array}{c}
a_T \\ \hline
a_B
\end{array} \right),
\]
where $ a_T $ represents $ \alpha_{0}, \ldots, \alpha_{k-1} $
and $ a_B $ 
represents
$ \alpha_{k}, \ldots, \alpha_{n-1} $.

We use the FLAME notation to indicate various objects~\cite{}. We use capital Roman letters to indicate matrices, (e.g. $A, B$), lower case Roman letters to indicate vectors (e.g. $a, b$), and lower case Greek letters to indicate scalars (e.g. $\alpha, \beta$). A matrix or a vector can split into submatrices and subvectors, which is indicated by subscripts (e.g. $a = \left( a_0 | a_1 | \dots | a_{N-1} \right)$.

While deriving algorithms, as we will see in the following sections, we focus on partitioning a vector into two parts $a = \FlaTwoByOne{a_T}{a_B}$. This partitioning helps separate out the part of the vector that has been  processed, and the rest of the vector that has not been processed yet. During the current iteration of the loop we expose an element of the vector from the partition that has not been processed yet, use it to update the current iteration and then merge it with the partition that has been processed.

As part of an algorithm, a ``current element'' 
(coefficient in our example) needs to be exposed that moves from the set of coefficients with which computation has not yet completed, $ a_T $ to the set 
of coefficients with which computation has completed, $ a_B $.
This is captured by repartitioning
}
\renewcommand{\operation}{}

\renewcommand{\routinename}{\operation}

\renewcommand{\precondition}{P_{\rm pre}}

\renewcommand{\postcondition}{P_{\rm post}}

\renewcommand{\invariant}{\PInv}

\renewcommand{\blocksize}{blank}

\renewcommand{\guard}{G}

\renewcommand{\partitionings}{$\phantom{something}$}

\renewcommand{\partitionsizes}{$ \phantom{ a_B  has  0  rows} $}

\renewcommand{\repartitionings}{
$
\phantom{repartitioning}   
$
}

\renewcommand{\repartitionsizes}{$\phantom{ a_B  has  0  rows}$}

\renewcommand{\moveboundaries}{
$
\phantom{move}   
$
}

\renewcommand{\beforeupdate}{$P_{\rm before}$}

\renewcommand{\afterupdate}{$P_{\rm after}$}

\renewcommand{\update}{$
\phantom{
\begin{array}{l}
    \psi := \alpha_1 + \psi \chi
\end{array}
}$
}
\subsection{The worksheet}

\begin{figure}[tb!]
    \begin{center}
    \FlaWorksheet
    \end{center}
    \caption{Blank worksheet used to systematically derive algorithms using the FLAME notation.}
    \label{fig:blankWorksheetFlame}
\end{figure}

Figure~\ref{fig:blankWorksheetFlame} shows a blank worksheet that assists in systematically deriving algorithms. The lines with a gray background will hold the commands of the algorithm and the lines with a white background will hold the proof of correctness. By systematically filling out the worksheet in the order of the step~number in the left column we can systematically derive algorithms.

\subsection{Deriving algorithms to evaluate polynomials}
We will work through the steps listed in the worksheet, and derive algorithms to evaluate polynomials, using the FLAME notation.
The completed worksheet is given in Figure~\ref{fig:HornersFlame}.
The reader may wish to print this to follow along.
This worksheet, using the FLAME notation, is what we call the FLAME methodology~\cite{FLAME,Recipe}.

\renewcommand{\precondition}{T}
\renewcommand{\postcondition}{\psi = \pi(a, \chi)}
\renewcommand{\invariant}{\psi = \pi\left(a_B, \chi \right)}
\renewcommand{\guard}{m(a_B) < m(a)}
\renewcommand{\partitionings}{$\psi:=0$ \\ $a \rightarrow \FlaTwoByOne{a_T}{a_B}$
}
\renewcommand{\partitionsizes}{	$ a_B $ has $ 0 $ elements}
\renewcommand{\repartitionings}{
	$  \FlaTwoByOne{a_T}{a_B}
	\rightarrow
	\FlaThreeByOneT{a_0}{\alpha_1}{a_2} 
	$
}
\renewcommand{\repartitionsizes}{
	$ \alpha_1 $ has $ 1 $ row}
	
\renewcommand{\moveboundaries}{
	$  \FlaTwoByOne{a_T}{a_B}
	\leftarrow
	\FlaThreeByOneB{a_0}{\alpha_1}{a_2}
	$
}
\renewcommand{\beforeupdate}{$\psi = \pi \left(a_2, \chi \right)$}
\renewcommand{\afterupdate}{$\psi = \alpha_1 + \pi\left(a_2, \chi \right) \chi$}

\renewcommand{\update}{
	$
	\begin{array}{l}          
    \psi := \alpha_1 + \psi \times \chi
	\end{array}               
	$
}

\begin{figure}[tb!]
   \begin{center}
    \FlaWorksheet
    \end{center}
    \caption{Derivation of Horner's Rule to evaluate polynomials.}
    \label{fig:HornersFlame}
\end{figure}

\subsubsection{Step~1: Precondition and postcondition.}
In Section~\ref{subsec:evalPolyIdx}, the precondition was $ 0 \leq n $.  In FLAME notation, this would translates to $ 0 \leq m( a ) $.  However, since the use of lower case letter $ a $ indicates $ a $ is a vector, and a vector must have nonzero length, the precondition now becomes $ T $ ({\em true}).
The postcondition is given by  $\postcondition$,
where 
$ \pi( a, \chi ) $ equals the polynomial defined by the elements of $ a $, evaluated at $ \chi $.
Implicit is the fact that the values in $ a $ and $ \chi $ do not change.

\begin{figure}[tb!]
\begin{center}
\begin{tabular}{|l I l l  | l l |} \hline
&
\multicolumn{2}{c|}{Invariant in traditional notation}
&
\multicolumn{2}{c|}{Invariant in FLAME notation}
\\ \whline
	Invariant 1 & $y = \sum_{i=0}^{k} a_i x^i $&$ \wedge 0 \leq k \leq n$
	&
	$\psi = \pi \left(a_T, \chi\right) $
	&
	\\\hline
	Invariant 2 & $y = \sum_{i=k}^{n-1} a_i x^i  
	$&$
	\wedge   0 \leq k \leq n$ 
	&
	$\psi = \pi \left(a_B, \chi\right) \chi^{m(a_T)}$ &
	\\\hline
	Invariant 3 & $y = \sum_{i=0}^{k} a_i x^i\,\wedge z= x^k\, $&$
	\wedge  \, 0 \leq k \leq n$
	&
	$\psi = \pi \left(a_T, \chi\right)$
	&
	$\wedge z= \chi^k$
	\\ \hline
	Invariant 4 & $y = \sum_{i=k}^{n-1} a_i x^i\,\wedge z= x^k\, $&$
	\wedge  \, 0 \leq k \leq n $
	&
	$\psi = \pi\left(a_B, \chi\right) \chi^{m(a_T)} $ &
	$\wedge z= \chi^k $\\ \hline
	Invariant 5 & $y = \sum_{i=k}^{n-1} a_i x^{i-k} $&$
	\wedge  \, 0 \leq k \leq n $
	&
	$\invariant$
	&\\ \hline
\end{tabular}
\end{center}
\caption{Five loop invariants for computing $ \pi( a, \chi ) $.
While we give the loop invariants in traditional notation for comparison, the loop invariants in FLAME notation are actually systematically derived from the recursive definition of $ \pi( a, \chi ) $ by identifying partial progress.}
\label{fig:LoopInvariantsFLAME}
\end{figure}

\subsubsection{Step~2: Determine the loop invariants.} To derive loop invariants for computing $ \psi = \pi( a, \chi )$, we use the fact that if $ a $ is partitioned into $ a_T $ and $ a_B $. Then 
\[
\psi = \pi\left( \FlaTwoByOne{a_T}{a_B}, \chi\right)
=
\pi\left(a_T, \chi\right) + \pi\left(a_B, \chi\right) \chi^{m(a_T)},
\]
where, as before, $ m( a_T ) $ equals the size of vector $ a_T $.
From this recursive definition of the operation, the five loop invariants 
tabulated in Figure~\ref{fig:LoopInvariantsFLAME} become evident, since the loop invariant must indicate partial progress towards the postcondition.
We focus on Invariant 5. The reader can repeat this process with the other invariants. 

{\em Loop invariants are systematically derived from the recursive definition of the operation.}

\subsubsection{Step~3: Determine the loop guard.} 
Instantiating the postcondition and the invariant, the bottom of the worksheet becomes 
\begin{center}
\begin{tabular}{| c | p{0.9\textwidth} |}\hline
2,3 & 
$ \left\{ 
\begin{minipage}{0.88\textwidth} 
$ \ShowInvariant \wedge \neg G $ 
\end{minipage}
\right\}
$
\\ \hline
1b & 
$ \left\{ 
\begin{minipage}{0.88\textwidth} 
$ \ShowPostcondition $ 
\end{minipage}
\right\}
$
\\ \hline
\end{tabular}
\end{center}

\noindent
which means that $\invariant \wedge \neg G $
must imply $ \postcondition$. 
 This suggests the guard $\guard$ since then $a_B$ is all of $a$ when the $\neg G$ is true%
 \footnote{$ m(a_B) > m(a) $ cannot happen since it is a subvector of $ a $}. 

{\em The postcondition and the chosen loop invariant dictate  the loop guard.}

\subsubsection{Step~4: Initializing the loop.}
At the top of the worksheet, given the precondition, we must determine the appropriate initialization such that we end in a state where the loop invariant is true. In other words, the Hoare triple $\{\precondition\} S \{\invariant \}$, must hold. This dictates that $ S $ be the  initialization $\psi:=0 $ and partitioning $ a \rightarrow \FlaTwoByOne{a_T}{a_B}$, where \partitionsizes.

{\em The precondition and the chosen loop invariant dictate  the initialization.}

\subsubsection{Step~5: Progressing through the loop.} To make progress towards the postcondition, at each iteration of the loop we expose one element of the vector $a_T$ that has not been processed yet, and at the bottom of the loop we include this element to the partition of the vector that has already been computed $a_B$. This  shown in Step~5a, and 5b in the worksheet.

{\em The initialization step and the loop guard dictate how to traverse the array.}

\subsubsection{Step~6: State after repartitioning.} At the top of the loop, the loop invariant must be true. The repartitioning Step~5a is merely an indexing step, exposing the ``next element.'' Since no computation occurs in an indexing step, we can express state of $ \psi $ in terms of the now exposed parts of $ a $.  This is a matter of recognizing that
\[
\pi( a_B, x ) = \pi( a_2,x )
\]
so that at Step~6 $ \psi = \pi( a_2, x ) $,
since $ a_B $ is renamed $ a_2 $.
This is entered as Step~6 in the worksheet.

{\em The loop invariant and the repartitioning of the array dictate the state in Step~6.}

\subsubsection{Step~7: State after computation.} 
Similarly, at the bottom of the loop, the loop invariant must be true. However, since we have made progress through the loop, the loop invariant must be true in terms of the exposed blocks in Step~5b. 
More formally, at Step~7
\begin{eqnarray*}
\lefteqn{\wp\left( \leftdq \mbox{\moveboundaries} \rightdq, \invariant \right)} \\
&=&
\pi( \FlaTwoByOneSingleLine{ \alpha_1 }{a_2}, \chi ) 
=
\pi( \alpha_1, \chi ) + 
\pi( a_2, \chi) \chi^{m( \alpha_1)}
=
\alpha_1 + \pi( a_2, \chi) \times \chi
\end{eqnarray*}
must hold.
This gives us Step~7, which is the state of the variables after the update in Step~8 is performed.

{\em The loop invariant together and the redefinition of what are the top and bottom part of the array dictate  Step~7.}

\subsubsection{Step~8: Update.}
The update in Step~8 must change the state of the variables from that given in Step~6 to that given in Step~7.
The update $ \psi := \alpha_1 + \psi \times \chi $ has that property.

{\em Step~6 and Step~7 dictate Step~8.}

\subsection{Discussion}

There is at least one notable difference between the worksheets in Figure~\ref{fig:blankWorksheetIndx} and Figure~\ref{fig:blankWorksheetFlame}.
\begin{itemize}
    \item 
In Figure~\ref{fig:blankWorksheetIndx},
we apply a backward analysis, systematically computing what the weakest precondition is from the bottom to the top of the loop body.
While this works well when the loop index is updated at the end of the loop body, the calculations of the weakest preconditions become cumbersome when the loop index is updated at the top of the loop.
\item
In contrast, Figure~\ref{fig:HornersFlame}
employs forward reasoning at the top to get from Step~2,3 to Step~6 along the lines of ``We know the loop invariant is {\em true} at the top of the loop and therefore in terms of the exposed parts $ \psi $ contains ... at Step~6.''  To derive Step~7, backward reasoning is employed:
``We know the loop invariant must again be  {\em true} at the bottom of the loop and therefore in terms of the exposed parts $ \psi $ must contain ... at Step~7.''
This reasoning applies regardless of whether the algorithm computes with elements of the vector from top to bottom or from bottom to the top.
\end{itemize}
The real shortcomings of the traditional notation becomes clearer when deriving algorithms for operations that involve matrices (2D arrays).  Details go beyond this paper.

\section{Cost evaluation of algorithms}

\newcommand{\CostInit}{$ C = 0 $}
\newcommand{\CostInvariant}{}
\newcommand{\CostPostCond}{}
\newcommand{\CostBefore}{}
\newcommand{\CostAfter}{}
\newcommand{\CostUpdate}{$ C := C + 2$}

\begin{figure}[tb!]
   \begin{center}
    \FlaCostWorksheet
    \end{center}
    \caption{Cost algorithm for Horner's Rule to evaluate polynomials.}
    \label{fig:HornersFlameCostOne}
\end{figure}

We now discuss a systematic way of analyzing the cost of 
an algorithm.
We will do so by focusing on the example from the last section, which evaluates a polynomial using Horner's Rule.  In computations like this, it is floating point operations (flops) that are considered ``useful work'' and are therefore usually counted.  One could easily choose to count other operations (like comparisons) as well.
In our domain, high-performance computing, an order of magnitude cost is not sufficient: we want the coefficient of the leading term in the cost function.

\subsection{Counting operations}

A simple way to start computing the cost is to count flops as they happen.  
This leads to the worksheet in Figure~\ref{fig:HornersFlameCostOne}, where we added this computation on the right-hand side of the worksheet.
The cost of the update $ \psi := \alpha_1 + \psi \times x $
is $ 2 $ flops (an add and a multiply).
Let us call this the {\em cost algorithm}.
If the algorithm were executed for a specific input, then upon completion we would find the cost in variable $ C $.

\subsection{Finding a closed-form expression}

One usually desires a closed-form expression of the cost, the cost function.
Letting $ C_k $ equal the cost after $ k $ iterations,
one recognizes that the cost algorithm defines a recurrence relation
\[
\left\{
\begin{array}{lcl}
C_0 & = & 0  \\
C_{k+1} & = & C_k + 2.
\end{array}
\right.
\]
Here $ k = m( a_B ) $, since our algorithm doesn't have a loop index.
Standard techniques taught in a discrete mathematics or algorithms course allows one to find a closed form expression:
\[
C_k = 2 k.
\]
or, in terms of the notation used for the algorithm,
$ C = 2 m( a_B ) $, before and after the execution of each iteration.  We will call this the {\em cost invariant}.
Since it will still be true after the loop completes, at which time $ m( a_B ) = m( a ) $, the cost of the algorithm is then $ 2 m( a ) $.

\subsection{A proof of the cost function}

\renewcommand{\CostInit}{$ C = 0 $}
\renewcommand{\CostInvariant}{$C = 2 m( a_B )$}

\renewcommand{\CostPostCond}{$C = 2 m( a )$}
\renewcommand{\CostBefore}{
$ C = 2m( a_2 ) $}
\renewcommand{\CostAfter}{
$C = 2m( \FlaTwoByOneSingleLine{ \alpha_1}{a_2} )
= 2(m( a_2) + 1)$}
\renewcommand{\CostUpdate}{$ C := C + 2$}

\begin{figure}[tb!]
   \begin{center}
    \FlaCostWorksheet
    \end{center}
    \caption{Proof of correctness of the cost function, $ 2 m( a ) $, for Horner's Rule.}
    \label{fig:HornersFlameCostTwo}
\end{figure}

We now recognize that the worksheet, annotated with the cost algorithm and the cost invariant, can be used to prove the correctness of the cost function, $ 2 m( a ) $,
as illustrated in Figure~\ref{fig:HornersFlameCostTwo}.
The usual proof by induction is now presented as a proof of correctness of the algorithm that computes the cost.
This reinforces the link between mathematical induction and how loops are proved correct.

\section{Impact and extensions}
\label{sec:enrichments}

We now briefly discuss how the  methodology has had practical impact as well as recent developments related to this work.

\subsection{Impact on algorithms for matrix operations}

The techniques described in Section~\ref{sec:FLAME} were developed as part of research on parallelizing (for distributed memory architectures) algorithms for dense linear algebra operations.  Key was the FLAME notation, first employed in a paper on inverting matrices~\cite{inverse-siam}.  This then led to a paper in which the notation was linked to the formal derivation of algorithms related to the solution of linear systems (LU factorization and solution of triangular matrices)~\cite{FLAME}. 
Also in that paper, and a subsequent paper~\cite{FLAME:API}, APIs were proposed for representing algorithms in code so that the correctness of the algorithm implied correctness of the implementation.  
The worksheet was introduced shortly after, in~\cite{Recipe}.  Since this early work, we have derived hundreds of algorithms which are incorporated in a high-performance linear algebra software library, {\tt libflame}~\cite{libflame_ref,CiSE09}.
A discussion of this work targeting the Formal Methods community can be found in a Formal Aspects of Computing paper~\cite{Bientinesi2013}, which uses the solution of a triangular Sylvester equation as an example.

\subsection{Correctness in the presence of roundoff error}

For numerical computations performed on computers, the use of floating point arithmetic inherently causes roundoff error to accumulate.  Correctness in this setting leads to the notion of backward error: ``Is the computed solution equal to a slightly 
perturbed problem?''~\cite{Wilkinson:1961:EAD:321075.321076,Higham:2002:ASN}.
The presented techniques can be used to systematically derive the backward error result for many algorithms, as shown in~\cite{Paolo:PhD,Bientinesi:2011:GMS:2078718.2078728}.

\subsection{Implementing the simple algorithm for computing families of algorithms}

In computer science, the purpose for exposing a systematic approach (algorithm) is usually so that it can be implemented in code.  This begs the question ``Can the computation of families of algorithms be implemented?''  The answer, for a large part of the domain of dense linear algebra computations, is ``Yes!'' as shown in~~\cite{Paolo:PhD,Fabregat-Traver2014:278,Fabregat-Traver2011:54,Fabregat-Traver2011:238},
culminating in an open source tool, Cl1ck%
\footnote{
Source code for Cl1ck available from \href{https://github.com/dfabregat/Cl1ck}{\url{https://github.com/dfabregat/Cl1ck}}.}.

Related is Microsoft's Dafny~\cite{co-induction-simply-automatic-co-inductive-proofs-in-a-program-verifier}, a language and program verifier that is not fully automatic but instead gives feedback about the formal correctness of programs to the user.

\subsection{How to choose the best algorithm}

``Best'' in our universe often means ``fastest.''
One strategy for finding the fastest algorithm is to derive and implement all,  and to then choose the one with the shortest execution time.
A good case study for this focuses on inversion of a symmetric positive-definite (SPD) matrix~\cite{Bientinesi:2008:FAR:1377603.1377606} on distributed memory architectures.  Another strategy is to leverage expert knowledge.

In yet another dissertation, Low~\cite{phd:low} shows that a property of the resulting algorithm can be an input to the derivation process.  In this case, he shows, the loop invariant for the algorithm with the desired property can be recognized from the recursive definition of the operation from which the invariant was derived.  This makes it only necessary to derive the corresponding algorithm with the described techniques.  In the dissertation, it is argued that this side-steps the so-called phase ordering problem for compilers.

\subsection{Scope}

The described algorithm for computing algorithms has been successfully applied in the area of dense linear algebra as well as to so-called Krylov subspace methods (iterative methods for solving systems of linear equations).  A question becomes ``How broadly applicable is it?''
In Chapter 9 of his dissertation, Low~\cite{phd:low} argues that the methodology in one form or another should apply to the class of Primitive Recursive Functions~\cite{PRF}.
Recently, it has been applied to graph operations that can be described with matrices~\cite{Lee:2017:FPC:3145344.3145484}.
It is our hope that others will extend the methodology to other domains.

\section{Conclusion}

In this paper, we have discussed a methodology (algorithm) for deriving loop-based algorithms.  The methodology is described as an eight step process that derives assertions and commands, thus yielding the algorithm hand in hand with its proof of correctness.  Key to conciseness and clarity is the FLAME notation, which hides intricate indexing details.
By presenting the methodology as a worksheet to be systematically filled, it better links concepts from Formal Methods to how  programmers (should) think as program.

At UT-Austin, the methodology has been taught to undergraduates in a class titled ``Programming for Correctness and Performance'' and it is also the topic of the Massive Open Online Course ``LAFF-On Programming for Correctness.''  This demonstrates that the methodology is simple enough to be taught to a broad audience. 

It is our opinion that all who write code should master this algorithm.

\section*{Acknowledgments}

This research was supported in part by NSF under grant numbers ACI-1550493, and ACI-1714091. We would like to thank our many collaborators who have contributed to this research over the years.

\bibliography{biblio,poly}	

\end{document}